\documentclass[10pt]{article}
\usepackage[dvips]{graphicx}
\begin{document}
\newcommand{\m}[1]{\ensuremath\mbox{\boldmath $#1$}}
\newcommand{\be}{\begin{equation}} \newcommand{\ee}{\end{equation}}
\newcommand{\ba}{\begin{eqnarray}} \newcommand{\ea}{\end{eqnarray}}
\newcommand{\nn}{\nonumber} \renewcommand{\bf}{\textbf}
\newcommand{\ra}{\rightarrow} \renewcommand{\c}{\cdot}
\renewcommand{\d}{\mathrm{d}} \newcommand{\diag}{\mathrm{diag}}
\renewcommand{\dim}{\mathrm{dim}} \newcommand{\D}{\mathrm{D}}
\newcommand{\integer}{\mathrm{integer}}
\newcommand{\LL}{\mathbf{\Lambda}} \newcommand{\R}{\mathbf{R}}
\renewcommand{\t}{\mathrm{t}} \newcommand{\T}{\mathbf{T}}
\newcommand{\V}{\mathbf{V}} \newcommand{\tr}{\mathrm{tr}}
\newcommand{\cA}{\cal A} \newcommand{\cB}{\cal B}
\newcommand{\cC}{\cal C} \newcommand{\cD}{\mathrm{\cal D}}
\newcommand{\cF}{\cal F} \newcommand{\cG}{\cal G}
\newcommand{\cL}{\cal L} \newcommand{\cO}{\cal O}
\newcommand{\cT}{\cal T} \newcommand{\cU}{\cal U}
\newcommand{\s}{\,\,\,} \renewcommand{\a}{\alpha}
\renewcommand{\b}{\beta} \newcommand{\e}{\mathrm{e}}
\newcommand{\eps}{\epsilon} \newcommand{\f}{\phi}
\newcommand{\fr}{\frac} \newcommand{\g}{\gamma} \newcommand{\h}{\hat}
\renewcommand{\i}{\mathrm{i}} \newcommand{\p}{\partial}
\newcommand{\w}{\wedge} \newcommand{\x}{\xi}
\input{epsf}

\title
{The Virgo Alignment Puzzle in Propagation of Radiation on Cosmological 
Scales}

\author{John P. Ralston$^1$ and Pankaj Jain$^2$\\
$^1$Department of Physics \& Astronomy,\\ Kansas University,\\
Lawrence, KS-66045, USA\\
email: ralston@ku.edu\\
$^2$Physics Department, I.I.T. Kanpur, India 208016 \\
email: pkjain@iitk.ac.in}
\maketitle \medskip

Abstract:  We reconsider analysis of data on the cosmic microwave 
background
on the largest angular scales.  Temperature multipoles of any order
factor naturally into a direct product of axial quantities and
cosets.  Striking coincidences exist among the axes associated with
the dipole, quadrupole, and octupole CMB moments.  These axes also
coincide well with two other axes independently determined from
polarizations at radio and optical frequencies propagating on
cosmological scales.  The five coincident axes indicate
physical correlation and anisotropic  properties of the cosmic medium
not predicted by the conventional Big Bang scenario.  We consider 
various mechanisms, including foreground corrections, as candidates for 
the observed correlations.  We also consider whether the propagation 
anomalies may be a signal of ``dark energy'' in the form of a condensed 
background field.
  Perhaps {\it light propagation} will prove to be an effective way to 
look for the
effects of {\it dark energy}.

\section{Introduction}

The cosmic microwave background is full of information.  It is
conventionally analyzed by expanding temperature correlations in 
spherical harmonics. Current literature is dominated by
attention to $l \sim 200 $.  The region of $l \sim 200 $ is predicted
by models of causal correlation at decoupling.  Multipoles of
small\footnote{Spin of the
quantum type cannot be confused, so we can use the simple term for 
brevity}
``spin'' $l$ contain information on scales too
large to possibly be physically correlated in the standard Big Bang
model. Yet it is well known that small $l$ quantities do not obey 
expectations, There has been a burst of
work\cite{LargescaleC} suggesting reinterpretation, usually 
with introduction of new parameters, in order to verify the Big Bang.

We maintain that the small-$l$ components play a scientifically pivotal
role.  They have the potential to falsify current hypotheses.
Falsification is more powerful than verification, so it is interesting
to re-investigate the low moments with an open mind.  We first set 
aside correlations, and use the temperature map $\Delta T (\theta, \, 
\phi)/T$ directly, because it has far more information.

The $l=1, \, m=0, \pm 1$ or {\it dipole} mode is commonly attributed to 
our motion
relative to the CMB rest frame.  This is a fine hypothesis, but we wish 
to think afresh about it.  We find it sobering to realize that the 
hypothesis has never been tested.  Everyone expects some component of 
dipole from a boost of our system, and yet the {\it sum} of any number 
of dipoles is a dipole in a
combined direction.  The annual motion of the Earth
about the Sun is an observable relative to a fixed dipole, and has
been measured.  However the balance of our proper motion is not an 
independent CMB observable.  Not unrelated, the confirmation of the 
``great attractor'' of local gravitational attraction correlated with 
the dipole seems to have gotten complicated: it remains an unfinished 
project.  The relative size of the dipole above an isotropic background 
might be
argued to confirm the orthodox interpretation.  Yet confirmation of 
three
$l=1$ variables with three free parameters is not a test.

If the dipole were the only observable, perhaps discussion would be 
moot. Yet for 30 years the attempt to resolve the relation with 
higher-spin
moments was frustrated by their unexpectedly {\it small} magnitude.  The
relationships between the low-$l$ moments are not the predicted ones
to this day.    It turns out that the low-$l$ multipoles contain 
additional group-theoretic information on directionality of the 
radiation background.   The dipole is simple because it is summarized 
by a
magnitude and a direction in space towards {\it Virgo}.  Presumably 
this was simply a random accident.   It is quite surprising that the  
various directions associated with higher multipoles
show mutual correlations in direction which are not predicted by the 
basic Big Bang.  This causes us to rethink the meaning of the low $l$
multipoles, preferring to work from the bottom-up as data guides us, 
and without deciding here what the origin of the effects may be.

Here is the plan of the paper.  In the next Section we develop a 
decomposition of the multipoles based on symmetry analysis.  There is a 
notion of invariantly preferred frames for any spin-$l$.   The axial 
quantities obtained coincide with those calculated independently by a 
numerical method.   We also review other possibilities, based on 
projecting arbitrary products of angular momenta representations into 
smaller dimension.    Section 3 reviews previously published work and 
compares it.  We separate statistical analysis into Subsection 3.4.  
This analysis is deliberately incomplete, and conservative, because the 
statistical significance of data depends upon the hypothesis.   Section 
4 discusses possible mechanisms for the new effects found.  The most 
ready explanation would perhaps be local effects on the CMB, but the 
importance of CMB data for cosmological deductions makes local 
perturbations hard to accept.   Moreover the existence of anisotropic 
correlations at optical frequencies, and independent radio frequency 
data, is not explained by any known model.   We find a loophole in 
current reasoning related to dark energy.  We suggest that propagation 
and polarization anomalies may be
an excellent probe of dark energy as compared to the more familiar 
Hubble flow probe of the energy-momentum tensor.  Perhaps {\it light } 
will prove to be an effective tool  to look for the
effects of {\it dark energy}.  Concluding remarks are given in Section 
5.

\section{Factoring Modes}

Let $|\psi_{l} >$ be the projection of the quantity $\psi(\theta, \, 
\phi)$ into angular momentum $l$. A moment is defined by
$\psi_{l m} =<\, l, m, \, | \psi(l) >$, where $|\, l, \, m >$ are
eigenstates of the angular momentum operators $\vec J^{2}, \, J_{z}$
in the spin-$l$ representation. The temperature measurements are real 
numbers, which produces a
constraint on the matrix elements: \ba \psi_{l m}=(-1)^{m}\psi_{l-m}
^{*}.  \ea

What are the natural ways to derive an axial quantity from a higher
multipole?  A simple assignment of a {\it vector} quantity is the
expectation \ba <\psi(l) | \, \vec J \, |\psi(l)>.  \label{spin1} \ea
This proposal has two flaws: (1) it is quadratic, not linear in the
data, and of doubtful physical interpretation.  The quantity in Eq.
\ref{spin1} is a map from representations $\psi(l) \otimes \psi^{*}(l)
\ra spin \, 1$ with dimensions of temperature-squared.  In the product
of two representations many new representations of high spin are
created and vectors of temperature-squared have no immediate meaning.
Perhaps for quadratic correlations the quadratic products of
temperatures giving spin-1 might mean something.  But we are
discussing the temperature.  (2) The quantity is zero.  The expectation
value in
 Eq.  \ref{spin1} vanishes for real representations.

     \begin{figure} \centering \epsfxsize=4.75in
\epsfysize=1.5in \epsfbox{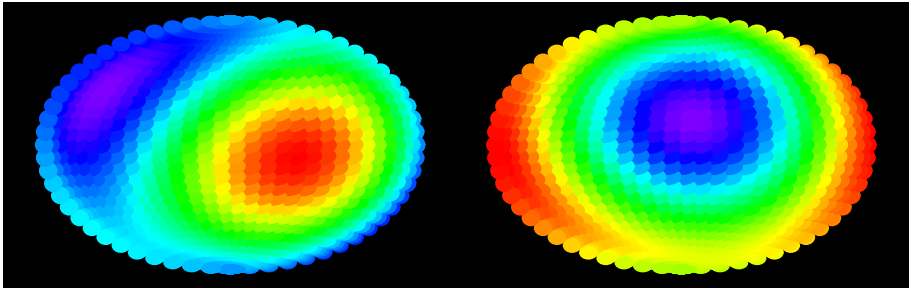} \caption{Aithoff-Hammer
projections of the {\it front hemisphere} $-\pi/2 < {\it l} < \pi/2$ of 
the CMB
temperature map in the spin $l=2$ quadrupole (left) and $l=3$ octupole
(right) modes.  The back side is trivial from the symmetry of the
modes.  } \label{QandOplot} \end{figure}

\subsection{A Linear Map}

Here we teach ourselves how to factor $l$-pole expansions.    The 
problem of factoring is
geometrical and would be
much the same for any angular quantity.

If the state $|\psi(l)>$ were an eigenstate of $J_{z}$ with eigenvalue
$m'$ in the frame where $J_{z}$ is diagonal, we could obtain a
relation $$ <m| \, J_{z}|\psi(l)> \, \ra m \, \delta_{mm'}.  $$ Here
$|m>=|l, \, m>$ is a concise notation.  Correspondingly, if we rotated
the operator $\vec J$, $$
\vec J \ra \vec J(\a, \, \b, \, \gamma) =U_{l} (\a, \, \b, \,
\gamma)\vec J U_{l} ^{\dagger} (\a, \, \b, \, \gamma), $$ then we
could determine the same relation for an eigenstate in an arbitrary
frame related by Euler angles $\a, \, \b, \, \gamma$.  It follows that
there is a linear map or ``wave function'' $\phi_{m}^{k}$ which
contains the information on frame-orientation: \ba \phi_{m}^{k}
=\frac{1}{\sqrt{l(l+1)}}<m| \, J^{k} |\psi(l)> .  \ea The ranges of
indices is $k=1\ldots 3$ and $m=-l \ldots l$.  We call this a wave
function in analogy with the Bethe-Salpeter wave function, which is
the matrix element of the operator creating a quantum state (analogous
to $J_{\pm} =J_{x} \pm i J_{y} $) and a fiducial vacuum state
(analogous to $|m=0>$.)  We are not doing quantum mechanics
and the physical observables exist at the level of this wave function,
not the wave function-squared.  Most importantly, $ \phi_{m}^{k}$ is
linear in the data and transforms like a vector in the index $k$ and
a rank $l$ tensor in the index $m$.

\subsubsection{Completeness}

The map to $\phi_{m} ^{k}$ is simply a coordinate transformation.
Start with $\psi_{m}= <m|\psi>$, of course for fixed $l$.  Make \ba
\phi_{m}^{k} &= & \Gamma_{mm'}^{k} \psi_{m'}; \nn \\
\Gamma_{mm'}^{k}&= & \frac{1}{\sqrt{l(l+1)}}<m|\, J^{k} \, |m'> .\ea
Repeated indices are summed.  Since $\phi_{m}^{k}$ has two indices,
and $\psi_{m'}$ has one index, the transformation is not obviously
invertible.  But the assignment of indices simply records the division
of the space.  The inverse of $\Gamma_{mm'}^{k}$ exists and is denoted
$\Gamma_{k}^{mm''}.  $ Construct the inverse directly: \ba
\Gamma_{k}^{mm''} = \frac{1}{\sqrt{l(l+1)}}<m|\, J^{k} \, |m''> ; \nn
\\
    \Gamma_{k}^{mm''} \Gamma_{m''m}^{k} &=&\frac{1}{l(l+1)}<m|\, J^{k} \,
    |m''> <m''|\, J^{k} \, |m'> , \nn \\ &=& \frac{1}{l(l+1)}<m| \, \vec
    J^{2} \, |m'>, \nn \\ &=& \delta_{mm'}.  \label{complete} \ea
    Technically $\Gamma_{k}^{mm''} $ is a pseudoinverse (Moore-Penrose
    inverse) for the two indices $m, \,k$ lying on the subspace of fixed
    $l$.  The map to $\phi$ is unitary for each $l$.

\subsubsection{Invariantly Preferred Frames}

Although $ \phi_{m}^{k}$ is neither symmetric nor Hermitian in
general, it has an invariantly preferred coordinate frame - in fact,
two frames.  The frames are found by the singular value decomposition
($svd$) \ba \phi_{m}^{k} =\sum_{\a}^{A} \, e_{k}^{\a}
\Lambda^{\a}u_{m}^{^{\dagger} \, \a}.  \label{svd}\ea The matrix
elements $e_{k}^{\a}$ and $u_{m}^{\a}$ are ``unitary'' (i.e.
orthogonal when real):\ba \sum_{m}\, u_{m}^{\a}(u_{m}^{\a'})^{*}=
\delta_{\a \a'} ; \label{umat} \\
\sum_{k}\, e_{k}^{\a}(e_{k}^{\a'})^{*}= \delta_{\a \a'} .  \ea It
follows that in a coordinate system rotated by $e$ on index $k$ and
rotated\footnote{``Rotation'' is used figuratively because $u$ lies on
$U(l)$ in the general case.  We did not investigate conditions for the
rotation on index $m$ to literally lie in an $SU(2)$ rotation
subgroup.} independently by $u^{\dagger} $ on index $m$ the expansion
of $\phi_{m}^{k} \ra (\Lambda^{1}, \, \Lambda^{2}, \ldots
\Lambda^{A})$ and {\it diagonal}.  This is what is meant by ``an
invariantly preferred coordinate system''.  In the preferred
coordinates, the frame $e_{k}^{\a} \ra \delta_{k \a}$, meaning that
$e^{1}\ra$ (1, 0, 0 \ldots), and so on.  The rest of $\phi_{m} ^{\a}$
becomes a definition of the particular vector chosen to be a fiducial
basis: the case $ \phi_{0} ^{\a}$ is a perfectly adequate ``vacuum.''
So the meaning of the $svd$ is that we can relate any state by a few
invariants and a rotation to an {\it invariantly determined} reference
frames.

Consistency requires that the number of terms $A$ does not exceed the
dimension of the smaller space.  For $l=1$ there is a degeneracy and
the number of terms $A=2 $.  In all cases $A \leq 3$.  The matrices
$u_{m}^{\a}$ are not a complete ``frame'' but have the nature of a
{\it coset}.  The eigenvalues $\Lambda^{\a}$ are invariant, real, and
by conventions of phase, positive.  The $svd$ is the direct
generalization to two independent spaces of the usual fact that a
Hermitian matrix can be diagonalized with the conjugate coordinate
transformations acting on the left and the right.  For those with
backgrounds only in quark physics, the $svd$ is the procedure to
reduce the CKM matrix to its invariant terms and unitarily-related
frames.  \paragraph{Degeneracy} If two or more eigenvalues
$\Lambda^{\a}$ are equal then $\psi$ is invariant under the
corresponding orthogonal symmetry.  While this is rare, a degeneracy
happens to occur in the data at hand, as seen momentarily.

\subsubsection{Axial Quantities}

Starting from our standard coordinate frame $\{ |i> \}$ with names
$i$, with $<j|i>=\delta_{ij  }   $, $<j|3> = (0, \,0, \, 1)$, and so 
on, the
three vectors $e_{k}^{1}, \, e_{k}^{2} , \,e_{k}^{3}$ represent three
preferred orthogonal axes in space.  Sometimes we will use the
notation $\hat e^{\a}$ for this reason.  We assess the importance of
these axes by the weights $\Lambda^{\a}$.  They parametrize the degree
of asymmetry.  Ordering the eigenvalues from largest to smallest, the
first frame element designated by $\a =1 $ describes a preferred axis.
We reiterate that eigenvectors describe an axis, in the meaning of an
oriented line in space, not signed-vectors in the sense of having a
direction.  In detail this comes because the signs of the $e^{a}$ are
conventions fixed by the signs of the weights $\Lambda^{\a}$.
Numerical code can conceal this fact.

\subsubsection{The Quadrupole}

The quadrupole is very instructive.  A quadrupole set of 5 numbers
$(q_{-2}, \, q_{-1}, \, \ldots q_{2})$ is just the relisting in the
angular momentum basis of the 5 elements of a traceless symmetric
3$\times $ 3 Cartesian tensor $Q_{ij}$: \ba Q_{ij}= V_{ij}^{m}
\psi_{m}(l).  \ea Here $V_{ij}^{m}$ is a unitary set of coefficients
to change indices $m$ to indices $ij$.

Such a tensor has an obvious preferred frame in which it is diagonal:
\ba Q =\sum_{\a}^{3} \, |q^{\a} >\Lambda_{Q} ^{a} <q^{\a} | .  \nn \ea
How are the obvious frame elements $|q^{\a} >$ related to the frame
elements $|e^{\a} >$ defined by $svd$?

They are the same frame, as evident by symmetry of diagonalizing on
the left.  The demonstration is awkward and we simply verified the
fact with numerical examples.

\subsection{Frames Defined by Expectation Value}

De Olviera {\it et al } (dOZTH)\cite{DO} have obtained a special axis
for the CMB quadrupole and octupole by the following procedure.  Define 
$\hat
n=\hat n(\theta, \, \phi)$ as a unit vector depending on angular
parameters $\theta, \, \phi$.  Find the maximum of the quadratic
expectation of $(\vec J \cdot \hat n)^{2} $, \ba <\psi(l)|\, ( \vec J
\cdot \hat n )^{2} \,| \psi(l)> \ra max.  \label{max} \ea   The sign
of $ \hat n$ is, of course, not determined by Eq. \ref{max}. dOZTH
report finding maximal unit vectors by a
numerical search with the 3,145,728 pixel map.  Using galactic
coordinates and the standard conventions of axes (see {\it
Appendix}), they find for the quadrupole and octupole: \ba \hat
n(l=2) = \pm(
-0.1145 , \, -0.5265 , \, 0.8424); \nn \\
\hat n (l=3) = \pm( -0.2578 , \, -0.4207 , \, 0.8698) .  \ea We think 
this
is an important result and we will return to it.

Besides one axis of absolute maximum weight, we observe there are
always two other $\hat n$ per multipole giving {\it local} maxima.
We see this by writing \ba <\psi(l)|\, (\vec J \cdot \hat n)^{2} \, |
\psi(l)> =\hat n_{k} K^{ kk'} \hat n_{k}; \nn \\
K^{ kk'} = <\psi(l)|\, J ^{ k} J ^{ k'} \,| \psi(l)>.  \ea Since the
commutator of the $J$'s gives no contribution, $K^{ kk'} $ is
symmetric in its indices.  By the Rayleigh-Ritz variational theorem
$\hat n$ will be one of {\it three} eigenvectors of $K$ when the
expectation is extremized.  (There would be $N$ such eigenvectors for
$N$ group generators.)  That is, the variational problem determines a
{\it frame}.

Earlier we mentioned that tensor quantities quadratic in the
temperature do not have the interpretation of temperature.  What is
the relation of the quadratically-produced $K^{kk'}$ axis and the one
from $svd$?  It is a matter of suppressing information in
$\phi_{m}^{k}$.  Consider $\phi_{m}^{k}$ as a matrix $\phi$ with
indices $k, \, m.$ Make a Hermitian square matrix that sums over the
information in index $m$: \ba \phi \phi^{\dagger}&=&\sum_{\a} \,
|e^{\a} >(\Lambda^{\a})^{2} <e^{\a} |;\nn \\ <k|\phi
\phi^{\dagger}|k'>&=& \sum_{m}<\psi(l)|\, J ^{ k} |m><m|J ^{ k'} \,|
\psi(l)>, \nn \\
&=&K^{kk'} .  \ea The $u$-frames disappear in the sum due to
unitarity.  This is a rather magical version of unitarity, as the
coset $u_{m}^{\a}$ with $-l<m<l$ indices only just spans the
three-dimensional subspace $\a=1\ldots 3$ it needs to span.  The
diagonal frame of $K$ coincides with the diagonal left-frame of $
\phi_{m} ^{k}$, namely the frame defined by the $e's$.   Meanwhile
the weights of $K$ are the weights of
$\phi$ squared.

We have shown, then, that a potential objection to an axis obtained 
from one
particular method quadratic in the data is not a worry, because it is
the same axis inherent linearly in the data.  The objection remains
true for other methods.

\subsection{Our Calculation}

We made our calculation (Table
\ref{tab:quadocto}) for the quadrupole and octupole by making the
spin-2 and spin-3 angular momentum matrices and diagonalizing by
$svd$.  The results are compatible with  dOZTH with small differences.  
We find
frame axes with largest eigenvalues:\ba \hat e^{(1)}(l=2)=( 0.112
,\, 0.506 ,\, -0.855) ;\:\:\:spin-2\nn\\
\hat e^{(1)}(l=3)=( 0.246 ,\, 0.399 ,\,-0.883).  \:\:\:spin-3\ \ea Our
sign conventions are fixed by $\Lambda^{\a }>0$, which is standard.
The eigenvalues and complete frames are listed in Table
\ref{tab:quadocto}.  We checked our result by comparing them to a
direct diagonalization of the traceless symmetric Cartesian tensor for
spin-2 , and by a third method using Clebsch-Gordan coefficients (see
below).  We believe the small differences are not significant: $\hat
n(l=2)\cdot e^{(1)}(l=2)$=0.9996.  There will be small differences
going from the pixels to the moments, and it is impossible to obtain
very high accuracy for an eigenvector by a variational method due to
the object function becoming stationary.

What about the eigenvalues?  Consulting Table \ref{tab:quadocto}, the
largest eigenvalue is indeed large, and apparently inconsistent  with an
uncorrelated isotropic  proposal.  We will discuss quantifying such 
features shortly.  The
octupole's two ``small'' eigenvales $\Lambda^{(2)}, \, \Lambda^{(3)}$
are nearly degenerate.   The frame of the octupole $e_{k}^{\a}(l=3)$
is far from isotropic but can be aligned with the other frames with
very small changes in the inputs.

\begin{table}
       \centering

    \begin{tabular}{ccccc}
\hline
\hline
   spin & $\Lambda^\a \:\:\:\:\:\:$ & $\hat e_1\:\:\:\: \:\:\:\:
\:\: $& $\hat e_2\:\:\:\: \:\:\:\: \:\: $ & $ \hat e_3 \:\:\:\: \:\:
\:\:\:\: $ \\ \end{tabular}

\begin{tabular}{ccccc}
       \hline
    l=2& 25.95 &  0.1115 &  0.5049 & -0.8559 \\
    & 17.70& 0.3874 & 0.771 & 0.5053 \\
     & 8.24& 0.9151 &-0.3880 & -0.1095 \\
\end{tabular}

\begin{tabular}{ccccc} \hline
      l=3 &47.35 & 0.2462 & 0.3992 & -0.8831 \\
       & 21.94 &- 0.9383 &  0.3262 & -0.1141 \\
       &  21.24  & 0.2425 & 0.8568 & 0.4550 \\
      \hline
      \hline

\end{tabular}
      \caption{ Cosmic Axial Coincidence of quadrupole and octupole
      orientation.  {\it Top Group:} Frame matrix elements from the
      quadrupole for $e_k^\a(l=2)$ in rows $\a=1..3$ and columns
      $k$=1...3.  {\it Bottom Group:} Same for the octupole
      $e_k^\a(l=3)$.  The octupole has a nearly perfect degeneracy,
      allowing the lower two rows to be freely rotated about the
      $e_k^{3}(l=3)$ axis.  In that case the two frames nearly coincide.
      }\label{tab:quadocto} \end{table}

\subsection{Other Methods}

We gave thought to the question of uniqueness, because it is
tedious to analyze what other axial quantities might be made using
higher powers of variables beyond the linear and quadratic
constructions discussed so far.

Under the rotation group, the product of two or more representations
$|l, \, m>  \otimes |l', \, m'>$ is reduced to a sum of tensors of rank
$|J, \, M>$ by the Clebsch-Gordan series or ``clebsches''.  So if we
want a quadrupole in order to make axes, and it must come from from
the product of two octupoles (say), then we need the $ 3 \otimes 3 \ra
2$ clebsches and it can be worked out.  We must distinguish the maps
of $|l, \, m> \otimes |l', \, m'>$ and $ |l, \, m>\otimes <', \, m'|$,
of course.

\subsubsection{Products of $l \otimes l^{*} $}

The clebsches for {\it real} symmetric projections $l \otimes l ^{*}
\ra 1$ vanish, just as the expectation $<\vec J>=0$ indicates.  In
fact $\vec J$ is just the operator of clebsches to make spin-1 from
such products.  We do not seek a vector, but a symmetric rank-2 tensor
to make axes.  Evidently $C^{ kk'} (2) = < J^{k} J^{k'}
-\frac{1}{3}\vec J^{2}\delta^{ kk'}>$ is the appropriate map from $l
\otimes l ^{*} \ra 2$, as just discussed.  The unit operator
$\frac{1}{3}\vec J^{2}\delta_{kk'}$ added to make a traceless
irreducible representation does not change the eigenvectors, but does
remove one independent eigenvalue.  The connection with clebsches
makes the procedure of dOZTH less miraculous and shows that the $svd$
frames, then, coincide with a Clebsh-Gordan series of squaring the
data, projecting down into spin 2, and finding the preferred frame of
the product.

    Although the result is trivial we checked our calculation of $\hat
    n_{2}, \hat n_{3}$ a third way by generating it by the CG series for
    $l \otimes l^{*}$.

\subsubsection{Products of $l \otimes l $}

What about the clebsch projections of products without complex
conjugation?  One finds that $l \otimes l \ra 1$ and $l \otimes l \ra
2$, and so on, are non-zero, and perfectly valid {\it complex}
quantities.  If one's experience is limited to neutrino physics, these
are like ``Majorana'' masses which take the non-conjugated products of
the Lorentz group and make a scalar.  From the $svd$ decomposition one
can gain insight on what they mean.  Form the outer product \ba
\phi_{k}^{m} \phi_{k'}^{m'}=\sum_{\a, \b} \, \, e_{k}^{\a}e_{k'}^{\b}
\Lambda^{\a}\Lambda^{\b} u_{m}^{^{\dagger} \, \a}u_{m'}^{^{\dagger} \,
\b}\ea Now a clebsch series to get spin-$J$ will project this by
contracting with a matrix $C_{mm'}^{J}$.  The overall form resulting
is \ba C_{mm'}^{J} \phi_{k}^{m} \phi_{k'}^{m'}= e_{k}^{\a}W_{J}
^{\a\b} e_{k'}^{\b}.  \label{impart} \ea This is far from a diagonal
form as there is no ``dagger'' on either symbol $e$.  The function of
$W_{J} ^{\a\b} $ is to mix the information in the $u_{m}
^{l}$ not transforming like vectors until spin-J comes out of the
mix.

While spin-2 can be made, it is complex, and it is hard to get excited
about an $m=1$ (say) term made from the product of $l=97, \, m=93$ and
$l=97, \, m=94, \, m=92$ that never existed before the data was
squared.  We do not pursue this or further products, given the 
invertibility
of $\phi_{m}^{k}$ (Eq.  \ref{complete}) that tells us it is complete.

\section{Cosmic Axial Coincidences}

We constructed symmetry methods to pursue a higher goal than 
classification. We turn to several striking
coincidences among the CMB\cite{DO} and other axial quantities, all
obtained from
radiation propagating on cosmological distances.  The coincidences pose 
a significant
puzzle.

First, the quadrupole and octupole axes coincide closely with each
other, and with a well-known third axis $\hat e_{dipole}^{1 }   $ 
pointed
toward {\it Virgo}.  The relations are:
\ba \hat e_{dipole}^{1} \cdot \hat e_{quad} ^{1} = 0.980 \nn \\
\hat e_{quad}^{1} \cdot \hat e_{octo}^{1} = 0.985 \nn \\
\hat e_{octo}^{1} \cdot \hat e_{dipole}^{1} = 0.939 .\ea

Compare these
to the default symmetry-based hypothesis that each vector is
uncorrelated with the others and distributed isotropically.  This
happens to be the basic Big Bang ($bBB$) {\it null}, and no further 
assumptions are needed
to employ it.  The coincidence of three vectors in nearly the same 
direction is not
at all expected.

\subsection{Coincidence with Independent Radio Frequency Observables}

In Ref. \cite{JR} an axis of anisotropy for radiation propagating on
cosmological scales was reported: in galactic coordinates the axis is 
found to be $$
\hat s_{JR} =(-0.0927,  0.4616, -0.8822). \label{JRaxis} $$ This axis
coincides with the
direction of {\it
Virgo}.  It is very remarkable that this axis coincides with axial
parameters from the CMB.

The result came from a test of propagation isotropy.  The paper
examined statistics of offset angles
of radio galaxy symmetry axes relative to their average polarization 
angles for
332 sources\cite{JR}.   The polarization offsets are variables
understood to be independent of intervening magnetic field.  Faraday
rotation measures $RM$ are removed source-by-source in linear fits to
wavelength dependence.  Further discussion is given
in the Section on {\it Mechanisms}.  Now there are {\bf {four}} 
coincident axes.

\subsection{Coincidence with Optical Frequency Observable}

In Refs. \cite{Huts} Hutsem\'{e}kers observed that optical polarization
data from cosmologically distant and widely separated quasars indicate
an improbable degree of coherence.  The alignment of polarizations on
a curved manifold (the sky) is tricky to assess and certain statistics
invariant under coordinate changes are needed.  In Ref.
\cite{JainNarainSarala} invariant statistics were constructed using
parallel-transport.  The facts of unusual coherence were confirmed,
and moreover a significant clustering of polarization coherence in
large patches in the sky was observed.

The axis of correlation obtained from a fit is
$$  \hat s_{Hut} =(0.0344,   0.4844, -0.8741)\ ,$$
in galactic coordinates. The axis given above has been obtained
by imposing the cut on the degree of polarization $p\leq 2$ \%.
This cut gave the largest optical polarization
alignment effect \cite{JainNarainSarala}. The axis obtained
by the full data agrees within errors.
This axis
coincides with the direction of {\it Virgo}.

Now there are {\bf {five}} coincident axes.

\subsection{CMB Polarization Observables }

COBE led the way in measuring moments of the temperature map beyond
the dipole.  WMAP \cite{WMAP1} observations yield highly detailed
information on CMB polarizations for the first time.  Among the
features not predicted by the $bBB$ is a highly significant
cross-correlation of the temperature and degree of polarization
(Stoke's $Q$) across very large angular scales.

WMAP uses an invariant polarization statistic equivalent to parallel
transport\cite{JainNarainSarala}.  The observed correlations are
inconsistent with
the $bBB$ predictions of uncorrelated isotropy on such scales.  A
subsequent re-ionizing phase has been proposed \cite{Kogut:2003et} 
to explain them.

We remarked earlier that the conventional $bBB$ has been ruled
out.  Insertion of a new hypothesis to explain the polarization
correlation justifies that remark.  Meanwhile we see no reason, even
under the burden of  the new hypothesis, for the polarization
quantities to have any preferred axial character.

For some reason WMAP has not released polarization data, as far as we
know.  We predict that WMAP may see an increased degree of polarization
in the direction of {\it Virgo}, or{\it  diametrically opposite}.
One cannot deduce this from the
correlation itself. We simply see a pattern here.   Our prediction is
empirical and  based on the fact that the CMB already points to {\it
Virgo} in the dipole, quadrupole, and octupole
axes.  Correlation with {\it Virgo} would simply be more of
the {\it Virgo Alignment}.

\subsubsection{Directionality}

Given the opportunity to speculate, the pattern of data also indicates
to us that CMB-derived polarizations should be aligned towards Virgo, 
along the
plane found by Hutsem\'{e}kers.  Like the others, this deduction does 
not
follow from the correlations published.  It follows on the basis of
inspecting patterns in data without imposing too many prior
assumptions.

\subsection{Quantification with Bayesian Priors}

Here we consider the statistical significance of some coincidences.

The null is uncorrelated isotropy.  The usual test of the null is \ba
P(null \, | \, data) = \frac{P(data \, | \, null) P(null)}{P(data)}.
\ea The probability of two axes $\hat e_{A}, \, \hat e_{B}$ coinciding
from the null to a degree equal or better than the data is the area of
the largest spherical cap touching them divided by $2 \pi$: \ba
P(e_{A} \cdot \hat e_{B} \, | \, null) =   1- |e_{A} \cdot \hat
e_{B}| .  \ea The probability of three coincidences is the
product: \ba P(e_{A} \cdot
\hat e_{B}, \, e_{B} \cdot \hat e_{C} \, | \, null) =    ( \, 1-
|e_{A} \cdot \hat e_{B}|  \, ) ( \, 1- |e_{B} \cdot \hat e_{C}|) \, ). 
\ea
We summarize the
numerical value of pairwise probabilities in Table \ref{tab:pvalues}.

\begin{figure}

       \centering

       \epsfxsize=5.2in \epsfysize=3.8in
\epsfbox{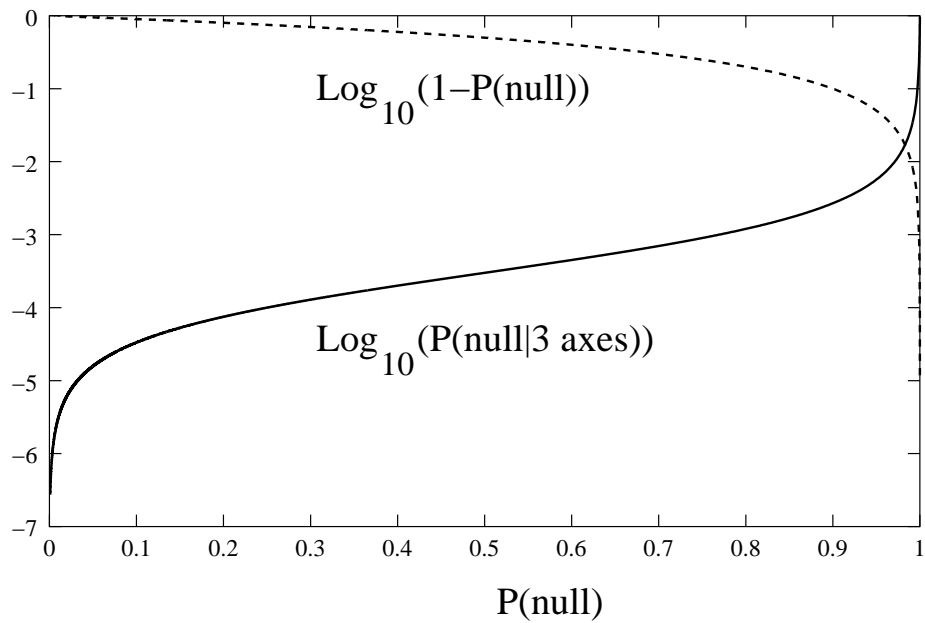}

\caption{Updated probabilities disfavoring the conventional
big-bang isotropic null hypothesis. The solid curve shows the
calculated probability of the hypothesis $log_{10}( \, P(null \, | \,
3 \, axes )\,) $ after seeing the
coincidence of dipole, quadrupole,
and octupole axes (only) as a function of prior prejudice $P(null)$.
The dashed curve shows $log_{10}(1-P(null))$, 
an estimator of the probability of 
the
alternate hypothesis.   
}
\label{fig:Prejudice}

\end{figure}

Continuing, we need $P(null)$, which comes from outside, and is
difficult to settle.  There may be an impulse to revert to large $l$
correlations and declare $P(null)=1$.  But large $l\geq 100 $ {\it
correlations} made in a manifestly isotropic way add no independent
information to the problem at hand.  The things commonly
predicted, namely magnitudes of the low $l$ moments, don't agree with
the standard null.  We find there is no good value for $P(null)$.

For argument let $P(null) =0.999$ simply because we can say it is so.
We need an alternate hypothesis $P(axes)$ to proceed.  This will be the
hypothesis that axes are well-correlated.  It may appear retrospective, 
but
follows from the non-CMB data (Section 3).  Quibbling is pointless 
because in a
two-option contest $P(axes) =1-P(null) =0.001 <<1 $ is very unlikely
anyway.   By definition $P(e_{A} \cdot \hat e_{B}, \, e_{B} \cdot \hat 
e_{C}
)\sim 1$ if axes tend to be correlated.  So long as $P(axes) $ is small
enough the details of $P(axes)$ will not change the evaluation of the
data.  Now calculate by Bayes Theorem: after seeing the coincidence of
the dipole, quadrupole, and octupole axes, the updated probability of
the null: \ba P(null \, | 3 \, axes) \sim \frac{P( 3 \, axes \, | \,
null)P(null)}{P( 3 \, axes \, | \, null)P(null)+P( 3 \, axes \, | \,
cor)P(cor)} , \nn \\ =\frac{0.02 \times 0.015 \times 0.999\, }{0.02
\times 0.015 \times 0.999\, + 1 \times 0.001} \nn \\
=0.23.  \ea The null survives because we used such a strong prejudice
favoring the $bBB$.  Yet if one had expressed less confidence, say
$P(null)=0.990$, the result would have changed, yielding $ P(null \, |
3 \, axes) \ra 0.029, $ and ruling out the null.  As a crisp summary,
{\it the null is ruled out with $97.1 \%$ confidence level, or indeed
much more, unless one's prior prejudice favoring it is $99.0 \%$ or
higher}.  For convenience we show parametrically the data-driven
confidence level in the null and the alternative hypothesis, as a 
function of
initial prejudice (Fig.  \ref{fig:Prejudice}).

\begin{table}
    \centering
    \begin{tabular}{c|ccccc}

        \hline
         & dipole & quad & octo & JR & Hutsem\'{e}kers \\
\hline dipole & & 0.020 & 0.061 & 0.042 & 0.024 \\
quad& & &0.015 & 0.023 & 0.004 \\
octo & & & & 0.059 & 0.026 \\ JR & & &
& & 0.008 \\
\hline
\end{tabular}
\caption{Probabilities as P-values $P$, for the axis coincidence
listed to equal or exceed that
seen in the data under an uncorrelated isotropic distribution.  By
ordinary convention $1-P$ is the confidence level that the
uncorrelated isotropic hypothesis is false.  }\label{tab:pvalues}
\end{table}

\subsubsection{Other Arguments}
It is interesting how a coincidence can be explained away by a
Bayesian re-arrangement of logic.  Assuming from the boost
interpretation that the dipole axis ``can be of no cosmological
significance'' (goes the argument), we remove one vector freedom in the
product of three possible probabilities.  Then only two things
coincide, and the astonishing improbability of the coincidence is
reduced to a merely alarming value.

Is this a swindle?  It depends.  Nobody can argue with a Bayesian
argument because the proponent who defines the priors may define their
own priors.  The simpler and more direct expression of prior prejudice
will hold that {\it all} the low $l$ moments have no (or little)
cosmological significance. In that case, nobody should not have even
looked at (let
alone paid millions for) the data!

There is a Bayesian response: let $P(junk)$ be the probability that
low-$l$ is junk.  ``Cosmic variance'', in the assertion that the
fluctuations of a few degrees of freedom tend to equal their value, is
for us on the logical track proportional to $P(junk)$.  In that case the
comparison of three axes would not be meaningful, except to update
prior beliefs in $P(junk)$.

We are not ready to assume that the bulk of the data in a large-scale
sense is just extremely expensive junk.  Given the cost and the
potential importance of the decision we feel it would be highly
irresponsible to do so.  The logical chain proportional to $1-P(junk)$
is the chain capable of testing the null.  The smallness of
$1-P(junk)$ is the entrance price for asking challenging questions.
One is not supposed to assess the entrance fee a second time when
something surprising is actually found.

Under the original hypothesis, and not adding further qualifications,
the conditional probability of coincidence $P(\hat n_{1}, \,\hat
n_{2}, \hat n_{3}, |\, bBB)$ of three axes is an extremely unlikely
occurence.

\section{Mechanisms}

The optical properties of cosmological propagation are not directly
connected to the Big Bang by general principles and need not have
significant other consequences.

It is also not necessary to have a mechanism to recognize regularities
in data.  Proposing mechanisms is helpful toward suggesting
interpretations and further tests.

Here we review information on the non-CMB axes and then turn to
potential mechanisms.

\subsection{Faraday Rotation Data}

\begin{figure}
\hskip -1.0in
\includegraphics[angle=-90,width=8in]{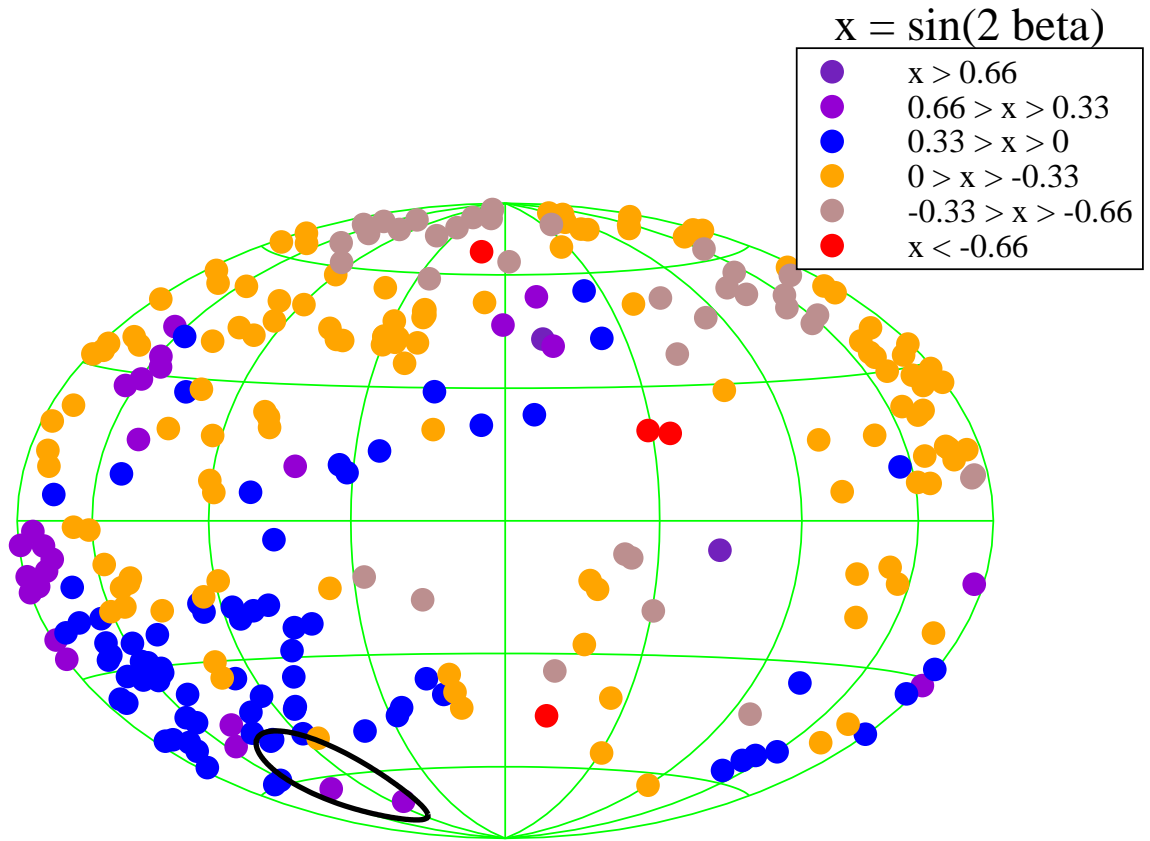}

   \caption{Covariant polarization offset statistics $x=sin(2 \beta)$,
where $\beta$ is the offset angle between galaxy symmetry axes and
polarization planes, for 265 radio galaxy sources mentioned in the
text.
The prefered axis $(l=101.36^o\pm 15^o,\ b=-61.91^o\pm 15^o)$
is shown as a black
cross. Axial clustering is readily visible to the eye. }

   \label{fig:FaradayFigure}
\end{figure}

There is a long history of observation of anisotropy at GHz
frequencies.  Controversies have been generated by discordant signals
in statistics initially supposed to represent the same phenomena.
Over a period of years, the discrepancies were traced to different {\it
symmetries} in the quantities being reported.  There is a consistent
signal in {\it odd-parity} statistics sensitive to a direction of
polarization rotation.

Birch\cite{Birch} observed patterns in data empirically.  Kendall and
Young\cite{KY} confirmed Birch with a likelihood analysis.  Bietenholz
and Kronberg\cite{BK} verified Birch's results, and confirmed a
statistically significant effect in a larger data set. They dismissed
the claim by turning to a statistic of {\it even} parity symmetry
developed by the statisticians
   Jupp and Mardia\cite{JuppMardia}.  Nodland and Ralston \cite{NR}
reported a redshift-dependent {\it
odd}-parity correlation in a set of 160 sources with known redshifts.
These data indicated correlations of offsets with an axis
oriented\footnote{The direction diametrically opposite to Virgo was
reported, but $\hat s_{NR}$ has no sign.}  toward {\it
Virgo}. Carrol and Field \cite{CF} dismissed the signal using RMS
angle measures that are coordinate-dependent\cite{JRsymclass}.
Eisenstein and Bunn re-plotted the data on the basis of {\it even}
parity and dismissed the signal by eye
\cite{Eisenstein:1997sa,ourResponse}. Loredo et al\cite{Loredo} found
no signal in an invariant statistic of {\it even } parity.

A symmetry analysis\cite{JRsymclass} showed that the Jupp-Mardia
($JM$) statistics come in even and odd-parity classes. Ref. \cite{JR}
reported axial correlations obtained from a larger set of 332 radio
galaxies.  Integrating over redshifts made statistics more robust and
increased the size of available data.  Both coordinate-free {\it
odd-parity} $JM$ statistics and likelihood analysis showed a strong
signal.  The case of highest significance, about 3.5 $\sigma$, was
found using an $RM$ cut parameter.\footnote{The statistical effect
of choosing an $RM$ cut parameter is of course included.  } The cut
is $(RM-\overline {RM}) \ge 6\ rad/m^2 $, where $\overline {RM} $
is the mean value of $RM$ in the sample.
A related study \cite{SaralaJain03}
yield axes consistent with the axis of Eq. \ref{JRaxis} within
errors. In fig. \ref{fig:FaradayFigure} we show the scatter plot of
$\sin(2\beta)$ for the 265 source data sample considered in Ref. 
\cite{JR}
after imposing the cut $(RM-\overline {RM}) \ge 6\ rad/m^2 $.
Here $\beta$ is the angle between the observation
polarization angle, after taking out the effect of the Faraday
Rotation, and the galaxy orientation angle. The figure shows the
averaged value of the variable $x = \sin(2\beta)$ at each position,
where the average is taken over the nearest neighbours which satisfy
$\hat r_1\cdot \hat r_2>0.95$.
Here $\hat r_1$ and $\hat r_2$ refer
to the angular positions of the two sources.

The mechanism for these correlations remains unknown.  The population
in $RM$ is peculiar, with a tall central spike and long tails. The
population and correlations might suggest an artifact of the
data-fitting procedure, which used the faulty concept of RMS angle.
However re-fitting the original data to obtain $RM$ and offset values
by an invariant
statistical method \cite{SaralaJain01} did not change the conclusions
significantly.

\subsubsection{Plasma Explanations?}

Perhaps Faraday rotation results could be explained away by the
ineluctable peculiarities of plasma electrodynamics. One study of
different statistics taken at higher frequencies saw no signal in a
selection of polarization and jet axes. \cite{Wardle}

Faraday rotation depends on the longitudinal component $B_{|| }$ of the
magnetic field, which must certainly fluctuate over cosmological 
distances.
However adiabatic propagation models traditionally used break down at 
field
direction reversals\cite{cork} $B_{|| } \ra 0$.  This conventional
physics was not included in the astronomical $RM$ analysis, and flips
polarizations.  It
might be a candidate for explanation.  Large angular scale
correlations in galactic or cosmological magnetic fields, however,
would be required for this mechanism to explain the Faraday rotation
data.  Galactic fields were always a concern for $RM$ measurements and
were thought to be understood.   We continue along these lines in the
subsection on {\it Foreground Mechanisms} below.

Cosmological magnetic fields of dipole, quadrupole or octupole
character have always formed a highly virulent threat to Big Bang
dynamics.  Little reliable information exists about cosmological
magnetic fields\cite{Zeldovich}, and data at frequencies dominated by
Faraday rotation
poses inherent difficulties of interpretation.

The CMB data, with frequency much larger than Faraday rotation
measurements, are thought to be comparatively free of such effects.
Compensations are made for the effects of the ionosphere and galaxy.
Under the standard hypotheses, the CMB measurements form an
independent observable which should not be related to Faraday rotation
data.

\subsection{Optical Data}

\begin{figure}
\epsfysize=2.5in
\hskip -1.0in
\includegraphics[angle=-90,width=8in]{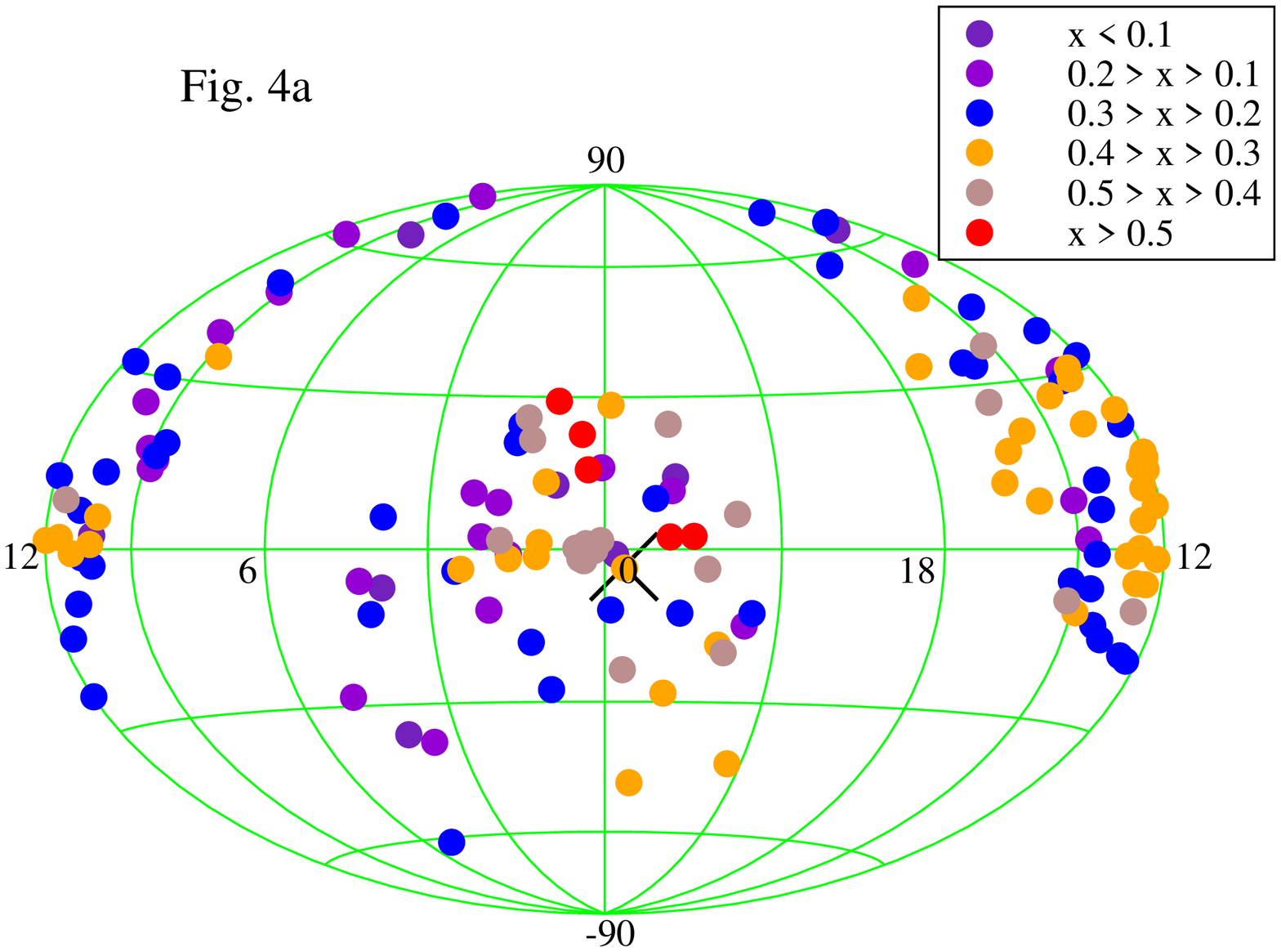}
\end{figure}
\begin{figure}
\epsfysize=2.5in

\hskip -1.0in
\includegraphics[angle=-90,width=8in]{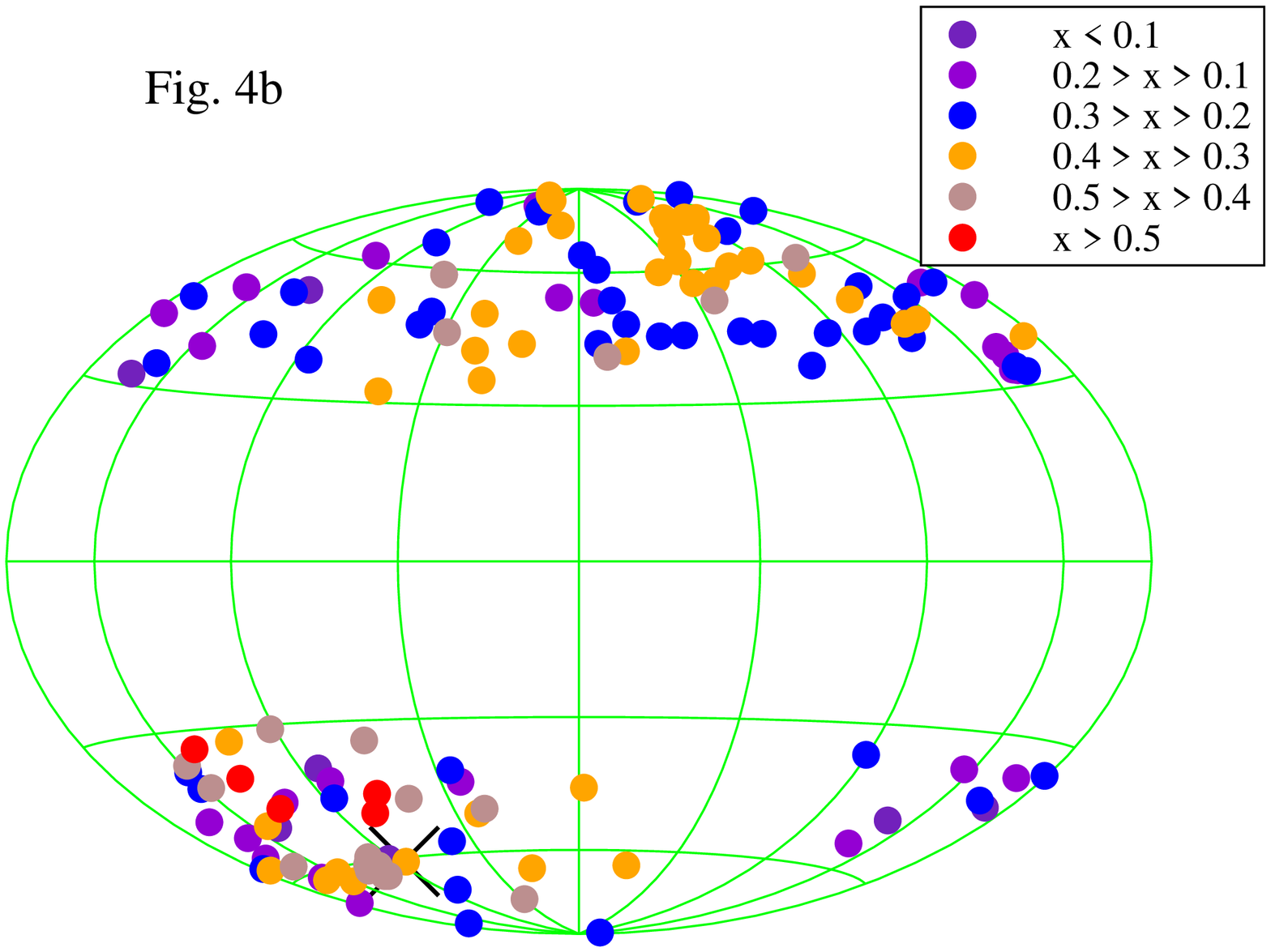}

   \caption{Optical polarization alignment observed by Hutsem\'{e}kers
   \cite{Huts} in (a) equatorial and (b) galactic coordinates.
   Here $x$ refers to the correlation statistic, defined in text, with
   the number of nearest neighbours chosen to be 28. The data within
   the galactic plane was deleted \cite{Huts} in order to minimize the 
effect of
   galactic extinction. Effects of population are taken into account in
   the construction of the prefered axis, shown as a black cross at
   $l= 85.94^o,\ b= -60.94^o$.
Axial clustering is readily visible to the eye. }

   \label{fig:OpticalFigure}
\end{figure}

The interpretation of Hutsem\'{e}kers result and subsequent verification
is complicated by the possible processes affecting optical
polarization along the trajectory.  The primary model to change the
degree of polarization is preferential extinction by dust.  However
any correlation of dust across vast cosmological distances would be as 
alarming
as a correlation of magnetic fields or the CMB radio spectrum.

Peculiarly, the data fraction of smallest polarizations has the
highest significance. In Fig. \ref{fig:OpticalFigure} we show the 
distribution
of correlation statistic $x=S_D^p$, defined in Ref. 
\cite{JainNarainSarala},
for the 146 source data sample after imposing
the cut $p\le 2$ \%. The statistic $S_D^p$ is the mean value of the 
dispersion
over the entire sample. A measure of the dispersion at each point is 
obtained
by maximizing the function,
$$ d_i(\theta) = {1\over n_v} \sum_{k=1}^{n_v} \cos\left[2\theta - 
2\left(
\theta_k + \Delta_{k\rightarrow i}\right) \right]\ ,$$
with respect to the angle $\theta$.
Here $(\theta_k + \Delta_{k\rightarrow i}) $
is the polarization at the position $k$ after parallel transport
to position $i$ along the great circle joining these two points.
The symbol $n_v$ is the number of nearest neighbours.
Fig. \ref{fig:OpticalFigure} uses $n_v=28$, a value giving
the strongest signal of alignment.

Although thought to be under control, we cannot rule out foreground
effects here and refer to the subsection on {\it Foreground Mechanisms}
below.

\subsection{Mechanisms}

We considered the following mechanisms:

\medskip

$\bullet$ {\it Foreground Mechanisms:} All of the data cited has
potential weaknesses in depending on models of the foreground, and in
particular, the effects of our Galaxy.

A great deal of ingenious technical effort has gone into taking this
into account without inducing biases.  It is unlikely that statistical
procedural errors exist.  Nevertheless there may be systematic {\it
physical} processes not taken into
account in current CMB foreground subtraction, which cause an
alignment with features of our Galaxy.  In the spirit of current CMB
fits, there is reason to believe that free adjustment of many
galactic magnetic field and optical extinction parameters might
explain not only the CMB data but also the non-CMB correlations.

It is somewhat hard to see how substantial foreground corrections that 
might make axial alignment correlations go away would not change CMB 
data and preserve its current interpretation as a pristine record of 
the early universe.

\medskip $\bullet$ {\it Metric Mechanisms:} The propagation of
polarizations is an old subject in General Relativity \cite{bire}.  In the
freely-falling frame where the metric is trivial, general covariance
says that the two polarizations propagate equally in Maxwell's theory,
and will preserve the state and direction of polarization.  This is
due to a symmetry of duality that the metric theory happens to
preserve.  Large polarization effects would need large curvature
effects that are not observed.  Given present information and beliefs
about the large-scale flatness of the Universe, we do not know of a
way for the large effects in the polarization data cited to be
explained.  \medskip

$\bullet$ {\it Magnetic Fields in the Early Universe:} The scaling
laws of magnetic fields are such that their imprint on the present
Universe can be hard to trace.  However large-scale magnetic fields
clearly could affect the $CMB$.  We cannot review all the work done in
this regard: It is simply hard to see how magnetic fields at small
redshift could
conspire in conventional physics to explain the Faraday offset
anisotropies, and it seems impossible to explain the Hutsem\'{e}kers 
data.

\medskip

$\bullet$ {\it Dark Energy:} If {\it dark energy} is due to a
scalar\footnote{As common in
cosmology we include pseudoscalars with scalars.} field there is a simple
generic mechanism to produce surprising optical effects that do
not necessarily disturb the other predictions of $bBB$ models.  Since
we have not seen this discussed elsewhere, we summarize it in broad
terms, referring the reader to a companion paper\cite{ourselves}.

During the process of expansion and decoupling, the photon field
changes dramatically over a wider range than any other particle.  The
change occurs in the {\it photon mass}.  The photon is massless {\it
en vacuo} due to the local nature of the theory and gauge invariance.
The photon is {\it not} massless in general, but propagates with a
mass $m_{p}^{2}$ given by the ``plasma frequency'' $\omega_{p}^{2}
=m_{p}^{2}c^{4} $.  The usual non-relativistic formula for the plasma
frequency is $$ \omega_{p}^{2} =\frac{e^{2}n }{4 \pi m_{e}}, $$ where
$-e$ is the electron's charge, $n$ is the density of light (electron)
carriers, and $m_{e}$ is the electron mass.  The dispersion relation
{\it en plasmo} is \ba \omega^{2}-c^{2} \vec k^{2}= m_{p}^{2}c^{4} .
\ea

Any scalar field $\phi$ coupling to the electromagnetic field strength
$F^{\mu \nu} $ will mix resonantly in a background magnetic field 
when the scalar mass is same as the
photon mass.  A typical Action is \ba S=\int d^{4}x \sqrt{g} \, \,
\left(\frac{-1}{4}F^{\mu \nu}F _{ \mu \nu} + g\phi \epsilon_{\mu \nu \a \b
}F^{\mu \nu}F^{\a \b}\right) ,\label{model}\ea which suffices to break
duality symmetry.  Parameter $g$ is a dimensionful coupling constant
that can be of exceedingly small size. 
The physical consequences of the mixing of this scalar field with
photons has been studied by several authors \cite{mixing}.
Since the plasma frequency and
the scalar mass are independent, and of different scaling
characteristics in general, it is hard to avoid scale-crossing.  In
particular, driving the photon mass to zero crosses every conceivable
scale at least once, to say nothing about re-ionization phases.  Once
duality symmetry is broken, polarization can be spontaneously
generated and also rotated under propagation.

No fundamental coupling is required because radiative corrections will
generate all couplings allowed by symmetry.  {\it The existence of a 
scalar
field --that is, dark energy -- is itself sufficient to guarantee that
the optical properties of propagation over vast distances and times
will not be those of the usual theory.}  Of course, the organization of
dark energy over huge scales is something that cannot be determined in
advance.  Models such as quintessence hardly encompass the
possibilities.

The observation of propagation anomalies is in fact a highly
conservative prediction of the current framework of cosmology, without
requiring new hypotheses.  While constantly cited, the data has ruled
out Einstein's early cosmological constant, the dark action \ba
S_{dark} =\int d^{4}x \, \sqrt{g} \, \Lambda .  \nn \ea Instead it is
necessary to have $\Lambda \ra \Lambda(\vec x, \, t)$.  
Now it would violate basic consistency requirements, such as causality,
if
what is currently meant by $\Lambda(\vec x, \, t)$ is not associated
with the modes of a field.
 We believe that what is meant by
$\Lambda(\vec x, \, t)$ is at least as dynamical as a field even more
than we believe in $\Lambda(\vec x, \, t)$

Unfortunately the current state of development of theory cannot relate
any particular field to any measurement to $\Lambda(\vec x, \, t)$.
One wonders why it is so important.  Meanwhile ordinary perturbative
methods can relate any field through radiative corrections to an
action such as Eq.  \ref{model}.  Thus propagation anomalies may be
an effective probe of dark energy, as compared to the rather indirect 
coupling to gravity via an energy-momentum tensor.  We reiterate that 
{\it light } propagation may yield a productive way to look for the
effects of {\it dark energy}.  \medskip

$\bullet$ {\it Dark Energy Plus Magnetic Fields}

Given that a resonant interaction inevitably occurs between dark
energy and electrodynamics, we have to admit ignorance on the
potential outcomes in terms of large scale magnetic fields.  This is
an extremely interesting dynamical problem. The interactions of
magnetic fields are not expected to retain isotropy in all features
of an expanding Universe.  Futher study along these directions
seems justified.

\section{Concluding Remarks}

There is nothing to fault in the beautiful technology
bringing moments of the CMB well above $l>100$.  Yet the small moments
   are problematic. The Virgo alignment appears to contradict the 
standard picture.

We believe that the tests possible in the near future and longer term
hinge on polarization quantities.  Perhaps the best way to search for
dark energy is with light.  WMAP may be poised to make a major
discovery that goes far beyond the verification of the models of
previous generations.

\section{Appendix: Conventions}
Here we collect details of conventions and procedures.

For spin $l$ we
define operators $$J_{z} = (-l, \, -l +1, \, ...  l-1, \, l) \:\:
(diagonal), $$ and we enter multipole moments as column vectors in the
same convention.  We define $$J_{x} = U(0, \, -\pi/2, \,0)
J_{z}U^{\dagger}(0, \, -\pi/2, \,0) \ ,$$ where the matrix elements of
$U(0, \,\beta, \,0)_{mm'}= d_{mm'}^{l}(\beta)$ are the Wigner
$d$-matrices with the standard
phase conventions.  Finally, $$ J_{y} =-i \ [ J_{z}, \, J_{x} \ ].  $$

The Hutsem\'{e}kers axis is obtained as follows.  Let $\hat n_{i}$ be
the unit vector coordinates of the $i$th source on the sky in
galactic coordinates.  Construct a two point comparison between the
$i$ and $j$ polarization, given by $$C(\hat n_i, \hat n_j ) =\cos[ \,
2\theta_i - 2
(\theta_j+ \Delta_{j\rightarrow i})]\ ,$$ where $(\theta_j+
\Delta_{j\rightarrow i})$ is the polarization angle obtained
after parallel transport of the polarization plane at position $j$ to
the position $i$ along the great circle connecting these two points.
Construct the
correlation tensor $T^{kk'}$,
$$  T^{kk'}  = \sum_{i, j}     \hat n_{i}^k  \hat n_{ j} ^{( k')}
   C(\hat n_i, \, \hat n_j )$$ By construction $T \ra 0$ in a
population of uncorrelated polarizations and positions.  Make the
matrix
$$ S = M^{-1/2}  T   M^{-1/2}$$
where $M$ is the matrix of sky locations
$$  M^{kk'} =  \sum_i    \hat n_{i }^k   \hat n_{i}^{k'}\ . $$  By
construction $S \ra <p>1$ in an isotropic sample with a net
polarization parameter $<p>$, so that $S \ra 0$ in the null.   The
Hutsem\'{e}kers   axis is the eigenvector of $S$ with the largest
eigenvalue.

\paragraph{Acknowledgements} Work supported in part under
Department of Energy grant number DE-FG02-04ER41308.

{ }

\end{document}